\documentclass[pra,reprint,aps,showpacs,merge, floatfix]{revtex4-1}
\usepackage{times,amsmath,graphicx,latexsym}

\begin{document}
\title{Superfluid Onset and Compressibility of $^4$He Films Adsorbed on Carbon Nanotubes}

\author{Emin Menachekanian}
\author{Vito Iaia}
\altaffiliation[Present address: ]{Department of Physics, Syracuse University, Syracuse, New York 13244-1130}
\author{Mingyu Fan}
\author{Jingjing Chen}
\altaffiliation[Present address: ]{Nankai University, Tianjin 300071, People's Republic of China} 
\author{Chaowei Hu}
\author{Ved Mittal}
\author{Gengming Liu}
\author{Raul Reyes}
\author{Fufang Wen}
\author{Gary A. Williams}
\affiliation{Department of Physics and Astronomy, University of California, Los Angeles, CA 90095}
\date{\today}  

\begin{abstract}
Third sound measurements of superfluid $^4$He thin films adsorbed on 10 nm diameter multiwall carbon nanotubes are used to  probe the superfluid onset temperature as a function of the film thickness, and to study the temperature dependence of the film compressibility.   The nanotubes provide a highly ordered carbon surface, with layer-by-layer growth of the adsorbed film as shown by oscillation peaks in the third sound velocity at the completion of the third, fourth, and fifth atomic layers, arising from oscillations in the compressibility.  In temperature sweeps the third sound velocity at very low temperatures is found to be linear with temperature, but oscillating between positive and negative slope depending on the film thickness.  Analysis shows that this can be attributed to a linearly decreasing compressibility of the film with temperature that appears to hold even near zero temperature.   The superfluid onset temperature is found to be linear in the film thickness, as predicted by the Kosterlitz-Thouless theory, but the slope is anomalous, a factor of three smaller than the predicted universal value.  
\end{abstract}

\maketitle

\section{Introduction}

 Carbon surfaces have been found to be ideal substrates to study the adsorption properties of noble gas atoms \cite{colebook}.  The hexagonal carbon rings form a highly ordered 2D atomic surface over distances that can exceed several hundred lattice constants, and this leads to layer-by-layer growth of the adsorbed films.  Studies \cite{bretz,greywall} with adsorbed $^4$He have shown a complex series of 2D phase transitions in the first two layers, between commensurate and incommensurate solids, and coexisting liquid and gas phases \cite{crowell,manousakis,manousakis1999,troyer,nyeki}.   The third layer is thought to primarily consist of gas-liquid coexistence, while the fourth and higher layers are assumed to be  predominantly liquid, now with the underlying first two layers solid \cite{zimmerli,crowell,campbell}.

 The onset of superfluidity in these films has shown unusual features.  Torsion oscillator measurements at low temperatures \cite{crowell} found superfluidity beginning at 1.85 layers, but it then disappeared on increasing the coverage to 1.95 layers, and in this region the superfluid density had an unusual logarithmic temperature dependence.  More recent measurements \cite{nyeki} have confirmed this reentrant behavior, which is quite different from the very rapid drop of the superfluid density to zero known to occur in the 2D Kosterlitz-Thouless (KT) transition \cite{kt,kosterlitz} in films on more disordered substrates such as glass \cite{rudnick1978} and Mylar \cite{bishop}.  It has been suggested that the superfluid transition in this region is not due to the excitation of vortex pairs, but arises from a density-wave ordering of low-energy roton-like excitations \cite{nyeki}.  On increasing the coverage the superfluidity then reappeared at 2.3 layers, exhibiting the characteristics of a finite-size broadened KT transition \cite{kotsubo,choprl}, and this was  tentatively ascribed to the gas-liquid coexistence which results in interconnected superfluid ``puddles" of finite radius \cite{crowell}.
 
It is only in the fourth layer that the superfluid transition begins to more closely resemble the characteristics of the KT transition seen on other substrates, with a sharp drop in the superfluid density at a critical temperature $T_{KT}$.  Crowell and Reppy \cite{crowell} compared their Grafoil torsion oscillator signal near 3.6 layers with the previous data on a Mylar substrate \cite{agnolet} that had nearly the same value of $T_{KT}$ (but a different film thickness), and concluded that the two substrates agreed ``reasonably well" with the KT theory.  However, it is unclear if such a comparison is entirely valid, since the surface of Mylar is extremely rough and disordered on atomic scales.  A more rigorous test of the KT theory is the predicted universal slope of the areal superfluid density $\sigma_s$ at $T_{KT}$ \cite{kt1977},
 \begin{equation}
\frac{{{\sigma _s}}}{{{T_{KT}}}} = \frac{{2{m^2}{k_B}}}{{\pi {\hbar ^2}}} = 3.49\,\frac{{{\text{ng}}}}{{{\text{c}}{{\text{m}}^{\text{2}}}{\text{K}}}}
\end{equation}
where $m$ is the helium atom mass.  As found previously \cite{rudnick1970,rudnick1974}, ${\sigma _s} = {\rho _s}(d - D)$ with ${\rho _s}$ the bulk superfluid density, $d$ the film thickness, and $D$ the effective nonsuperfluid "dead layer" thickness that defines the $T = 0$ onset thickness.  As noted below we find little evidence of normal fluid excitations below 1 K other than vortex pairs, so ${\rho _s} = \rho$, the bulk helium density, and taking one layer to be 3.58 \AA\, gives the equivalent slope
\begin{equation}
\frac{{{T_{KT}}}}{{(d - D)}} = 1.488\,\,{\text{K/layer}}\quad .
\end{equation} 

We have tested this relation using third sound measurements of $^4$He films adsorbed on multiwall carbon nanotubes, and find that the measured slope is a factor of more than three times smaller than the universal KT prediction over a thickness range of 3 to 4.4 atomic layers.  We have also found an unusual linear behavior of the third sound velocity with temperature at very low temperatures, with a slope that varies between positive and negative values as the film thickness varies, with a periodicity of  atomic layers.  Analysis shows that this can be ascribed to a linear decrease of the compressibility of the film with temperature, an unusual stiffening of the film with increasing temperature. 

\section{Experiment}
 \begin{figure}[t]
\begin{center}
\includegraphics[width=1.00\linewidth]{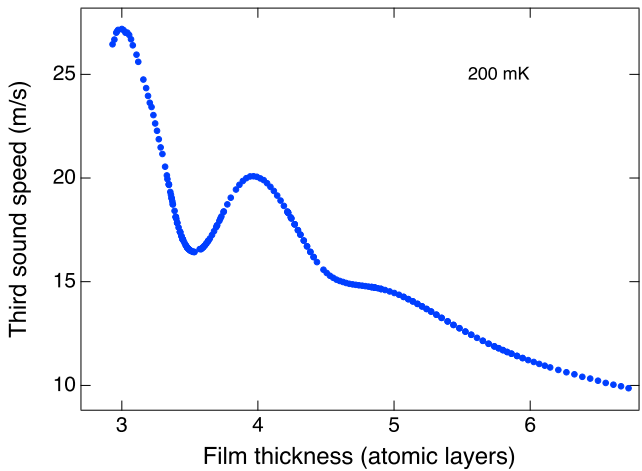}
\end{center}
\caption{(Color online) Third sound speed versus film thickness near 200 mK.}
\label{fig1}
\end{figure}

Third sound is a thickness wave in the film, where since the normal fluid is viscously locked to the substrate, only the superfluid component oscillates, giving rise to an accompanying temperature oscillation.   The Van der Waals interaction between the substrate and the helium provides the restoring force, and for an incompressible film the speed of the wave is given by \cite{rudnick1970,rudnick1974}
\begin{equation}
c_3^2 = \frac{{{\rho _s}}}{\rho }\frac{{3\,\gamma \,{\kern 1pt} (d - D)}}{{{d^4}}}
\end{equation}
where for a carbon substrate the Van der Waals constant $\gamma$ is taken as $44$ K(layers)$^3$ \cite{zimmerli}.  On a cylindrical substrate the value of $\gamma$ is reduced by the curvature, but a calculation \cite{emin2009} shows that for our diameter scales of 11-12 nm the reduction is less than 1\%, so we ignore such corrections.

The nanotubes used in these experiments are a commercial sample of multiwall tubes with a very tight distribution of diameters (as reported by the manufacturer \cite{southwest}), 10$\pm$1 nm, average length 2 $\mu$m, and average inner diameter 3 nm.  The tight distribution actually did not turn out to be terribly significant, as we found rather similar results \cite{emin2014} with a previous nanotube sample containing a range of diameters between 8 and 30 nm.  0.69 grams of the nanotube powder are lightly packed into a Plexiglass annulus of 3.5 cm mean diameter, 1.1 cm width, and 5 mm depth, pressed with a Plexiglass cover containing four bolometers at 90$^{\circ}$ spacing.  The plastic container became necessary when preliminary results showed that contact between the nanotubes and copper walls degraded the temperature oscillation of the third sound.  The plastic did not particularly impair thermal contact with the adsorbed films, as monitoring the sound velocity versus time after a typical temperature step of 30 mK showed equilibrium was achieved in less than 15 minutes for the thinnest films at the lowest temperatures, which were the slowest to change.  The bolometers are current-biased 200 $\Omega$ Allen-Bradley resistors (from the 1960's) with the case half-sanded off to expose the element; these get to hundreds of M$\Omega$ below 1 K.  The current bias was kept below 0.1 $\mu$A.  Since the nanotubes remained semiconducting below 1 K, it was necessary to interpose a Nuclepore filter between the nanotubes and the bolometer.  

It was found that a heater wire in the nanotube powder could not excite the third sound substantially without excessive dc heating, and hence a mechanical vibration source was attached to the outside of the copper cell sealing the nanotube assembly: a superconducting coil positioned about 2 mm from a neodymium permanent magnet attached to a copper 1 K shield.  A disadvantage of the mechanical excitation was that it also shook the dilution refrigerator attached to the cell, leading to spurious other resonances and limiting the maximum drive amplitude to the point where the refrigerator began to heat up.  A power-amplified ac signal was applied to the coil to generate the fundamental mode of the annular resonator (150-300 Hz) where one wavelength fits into the mean circumference of the annulus.  Slow frequency sweeps or passband-filtered noise drive voltages were employed in the vicinity of the resonances, with FFT analysis of the bolometer signals.  The FFT frequency resolution was typically 0.05 Hz, allowing the measurement of quality factors Q up to maximum values of about 3000, where the Q factor is the resonant frequency divided by the full width at half maximum of the power spectra resonances.  Such high Q's were only observed for thicker films in the fifth layer and above, giving signal-to-noise ratios greater than 100, but the Q factor and signal magnitude were strongly degraded in the thinnest films near the third-layer completion, requiring signal-averaging times of 40-60 minutes in this regime.  We could not observe any third sound signals for coverages below 2.93 layers where the Q factor of the sound resonance became too small (Q less than approximately 20-30) to observe any signal even at very low temperatures.  The fundamental resonance was actually a split resonance, split by as much as 16 Hz (proportional to the center frequency) that probably arose from a crack or defect in the powder packing, leading to different reflection probabilities for the counter-propagating waves of the mode.  The measured third sound velocities shown in Fig.\,1 are derived from the median frequencies.

\section{Results}

Figure 1 shows the measured third sound speeds as a function of the helium film thickness at temperatures within about $\pm$30 mK of 200 mK.  The layer completion points for the third, fourth, and fifth layers are readily evident by the maxima in the sound velocity, and we have defined the thickness scale for these points to be separated by 1.0 layers as expected theoretically \cite{Cheng1993}.  The intermediate points are scaled by the measured increments of added gas, with a linear scaling increasing by about 5\% due to the increasing surface area.  We also checked that the fourth layer completion point agreed to within 10\% of a prediction based on the surface area of the nanotubes (350 m$^2$/g) provided by the manufacturer \cite{southwest}.  Previous third sound measurements on ordered surfaces have seen similar oscillations in the sound speed \cite{zimmerli,mochel}, arising from the finite compressibility of the film.   Our speeds are smaller than those seen on a highly oriented pyrolitic graphite (HOPG) surface \cite{zimmerli} by a factor of about 1.7, due to an increased index of refraction $n$ for this nanotube substrate (a third sound speed lower than that on a flat substrate, $c_3 = c_{3,flat}/n$ due to geometry-dependent multiple scattering \cite{bernard}).  We believe our actual index of refraction is closer to $n = 2.0$, as discussed further in this paper, since HOPG surfaces typically have micron-sized crystallites that will give an index greater than 1.  An early experiment \cite{maynard} on HOPG gave results that can be interpreted as an index of order 1.3. 
\begin{figure}[t]
\begin{center}
\includegraphics[width=0.96\linewidth]{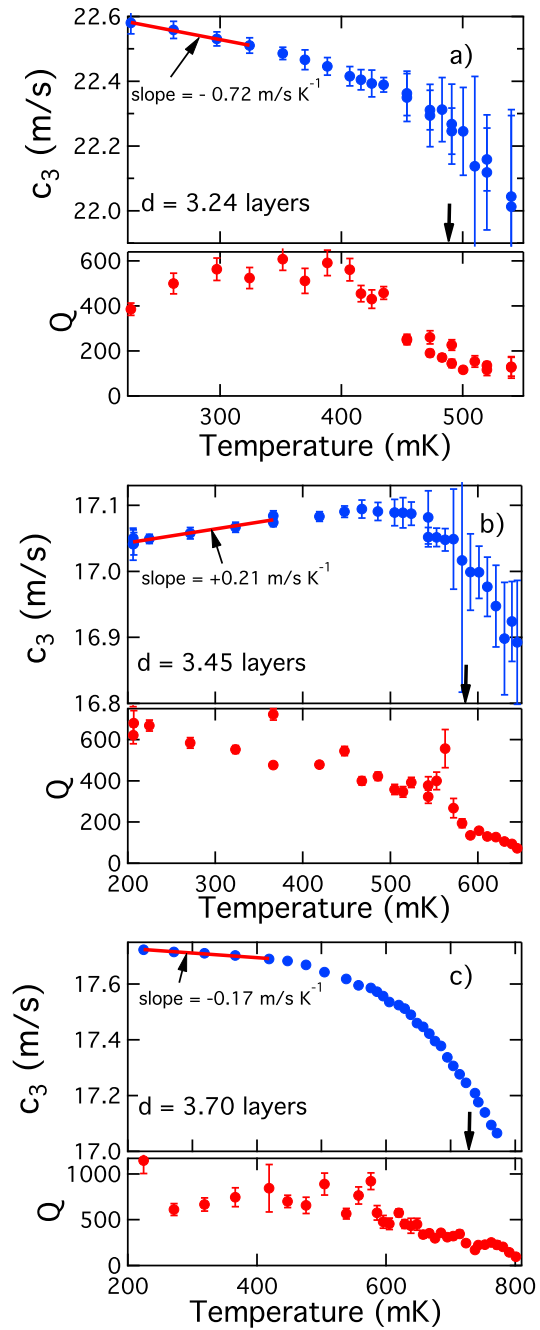}
\end{center}
\caption{(Color online) Temperature sweeps for different film thicknesses.  Arrows indicate the location of $T_{KT}$, where the Q factor falls below 200.}
\label{fig2}
\end{figure}

Temperature sweeps have been carried out at a number of fixed film thicknesses between 3 and 4.4 layers.  A sample of three such sweeps is shown in Fig.\,2, where the sound velocity and the Q factors are shown versus temperature.  At low temperatures a linear variation of the frequency with temperature is found, with the unusual property that the linear slope oscillates between negative and positive values with increasing film thickness.  The observation of the positive slope in Fig.\,2(b) near the half-layer fill (and also seen in our previous nanotube sample \cite{emin2014}) shows that this is not due to a normal-fluid excitation, since that could only decrease the superfluid density, and hence the third sound velocity.  Although our original thought was that this behavior might reflect an oscillating thickness change in the film (an oscillating expansion coefficient), we now believe instead that this is related to an unusual temperature dependence of the compressibility, to be discussed further below.  

The temperature sweeps of Fig.\,2 show a marked decrease at high temperatures in both the resonant frequency and the Q factor.  We believe this is the onset of the Kosterlitz-Thouless transition where vortex-pair excitations begin to reduce the superfluid density.  There is not a sharp drop in these quantities, which is likely due to finite-size effects caused by the 10 nm nanotube diameter.   The theory of the KT transition on a cylindrical surface has been formulated by Machta and Guyer \cite{Machta1989}, who find that vortex pairs of separation greater than the cylinder diameter become energetically unfavorable.  This leads to a finite-size broadening of the transition, a slower decrease of the superfluid density above $T_{KT}$ rather than the sharp drop at $T_{KT}$ seen on flat substrates.  Such finite-size effects have been seen experimentally with the KT transition of $^4$He films adsorbed on porous aluminum oxide, where the pore size is the confining length \cite{kotsubo,choprl,cho3}.  In measurements on a sample with pore sizes of order 10 nm \cite{choprl,cho3} the value of $T_{KT}$ as determined by the universal line of Eq. (2) coincided with the onset of a decrease in the superfluid density, but the drop to zero was considerably broadened compared to that on flat substrates.  In the current measurements the third sound is rapidly attenuated by the dissipation of the KT vortex pairs, and so although the onset of the transition is readily observable, only a small fraction of the total drop in $\sigma_s$ can be observed. 

Machta and Guyer \cite{Machta1989} also calculated the frictional dissipation arising from the motion of the vortex cores in response to the finite-frequency superfluid currents of the third sound on cylinders, and this turned out to be different from that of the porous material case.  In porous materials the energy landscape seen by the vortices is complex, and there are strong energy barriers that restrict the current-induced motion of the vortex pairs to stay within the pore size.  The limitation of the motion greatly lowers the dissipation of third sound, allowing the decrease of the superfluid density to be observed experimentally \cite{kotsubo} over nearly 90\% of the drop.  For the cylindrical geometry, however the barrier for counter-rotation of the vortices around the cylinder is quite small, and in a cycle of the motion the vortices can traverse net distances substantially longer than the cylinder perimeter.  Such motion gives rise to considerable dissipation, though still less than that observed on flat substrates.  It can be seen in Fig.\,2 that we are only able to observe the decrease in superfluid density of a maximum of about 10\% of the drop before the Q value becomes so low that we can no longer observe the signal.  For all of the data in Fig.\,2 the attempts to observe the signal at 10 mK above the highest-temperature data point were unsuccessful, even with hour-long averaging times.  Our observations are in qualitative agreement with the Machta-Guyer theory, but we are unable to carry out a quantitative check due to the rapid loss of signal right at the start of the transition.
\begin{figure}[t]
\begin{center}
\includegraphics[width=1.00\linewidth]{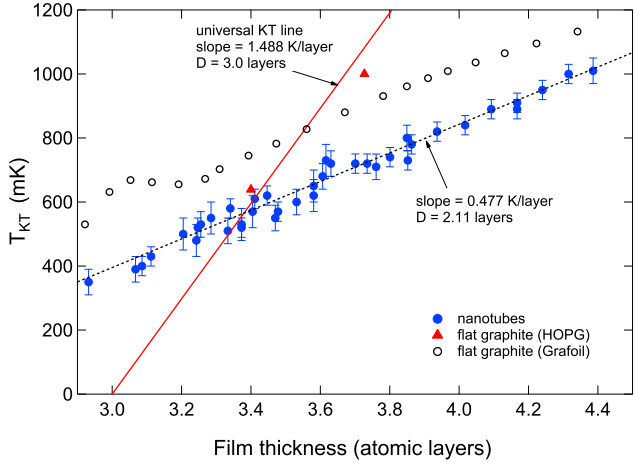}
\end{center}
\caption{(Color online) Superfluid onset temperatures versus film thickness. Filled circles:\,our data, open circles:\,Ref.\,[4], triangles:\,Ref.\,[9].  Dotted line is a fit to our data, while the solid line is the KT prediction of Eq.\,[2] with $D$ = 3.}
\label{fig4}
\end{figure}

To identify the onset temperature $T_{KT}$ for these broadened transitions we use as a criterion the point where the Q value drops below a value of about 200.  Since there are fluctuations in the Q, this leads to fairly large uncertainties in the value of $T_{KT}$, of order 30-50 mK.   However, since we are able to span a relatively wide range of film thicknesses between 3 and 4.4 layers (the upper limit where $T_{KT}$ approaches 1 K and the films begin to evaporate), the slope of the onset temperature is quite well defined, shown in Fig.\,3.  We find a best-fit slope of 0.477 K/layer, a factor of three less than the KT prediction, and an onset thickness of $D$ = 2.11 layers.  Even if we had adopted the flat-substrate criterion of total loss of the third sound signal \cite{rudnick1978}, the results would be completely similar, with the values of $T_{KT}$ only shifted upwards by perhaps 50 mK.  We found that our results for the KT onset in fact are quite similar to the flat Grafoil measurements of Crowell and Reppy \cite{crowell} in the same range of film thickness (though they did not note the disagreement with the KT slope in their paper).  We have plotted in Fig.\,3 their results for the torsion oscillator dissipation maximum, which should be slightly above $T_{KT}$, but scale with it.  Our value of $D$ is also consistent with their observation of superfluidity starting near  2.3 layers.  The onset measurements on HOPG \cite{zimmerli} did claim agreement with the KT line with $D$ = 3 layers, but with only two data points, shown in Fig.\,3.  The accuracy of this data is unclear, since both our results and those of \textcite{crowell} very clearly show superfluidity below 3 layers.

\begin{figure}[t]
\begin{center}
\includegraphics[width=1.00\linewidth]{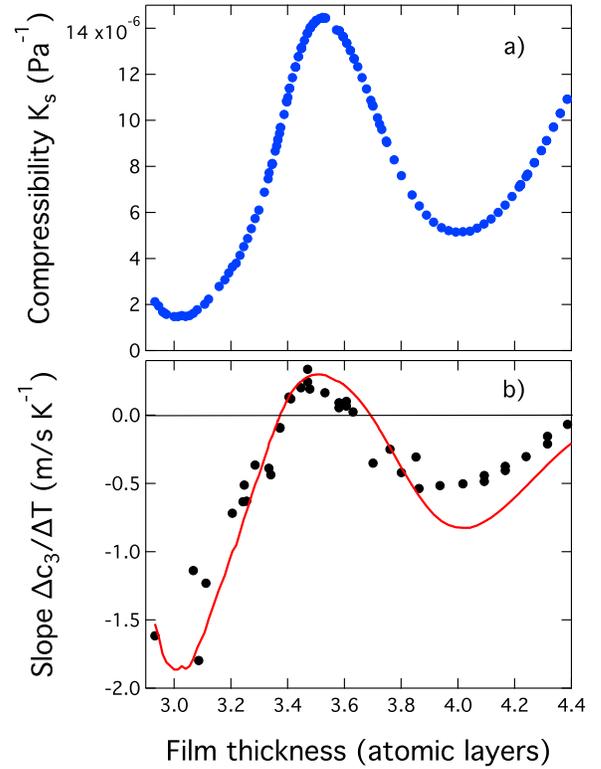}
\end{center}
\caption{(Color online) a) Compressibility versus film thickness at 200 mK, extracted from the $c_3$ data of Fig.\,1; b) Low temperature slopes of the third sound velocity, as in Fig.\,2.  The solid line comes from assuming a linear decrease in the compressibility with temperature, as discussed in the text.}
\label{fig3}
\end{figure}

The compressibility of the film can be computed from the third sound data following the calculation of Puff and Dash \cite{puff}.  The speed ${c'_3}$ including the compressibility can be written in terms of the first sound velocity $c_1$ as
\begin{equation}
c'^2_3 = {{c_3^2} \mathord{\left/
 {\vphantom {{c_3^2} {\left( {1 + \frac{\rho }{{{\rho _s}}}\frac{{c_3^2}}{{c_1^2}}} \right)}}} \right.
 \kern-\nulldelimiterspace} {\left( {1 + \frac{\rho }{{{\rho _s}}}\frac{{c_3^2}}{{c_1^2}}} \right)}}
\end{equation}
where $c_3$ is given by Eq.(3) and the isentropic compressibility is ${K_s} = {1 \mathord{\left/
 {\vphantom {1 {\rho {\kern 1pt} c}}} \right.
 \kern-\nulldelimiterspace} {\rho {\kern 1pt} c}}_1^2$.  Using our data for ${c'_3}$ with an index of refraction of 2.0, the extracted values of ${K_s}$ at 200 mK are shown in Fig.\,4(a).  The compressibility oscillates with film thickness, with minima at layer completions and maxima at half-layer points.  This is the same behavior seen in Refs.\,[4] and [9], though there a 2D definition of compressibility is used to derive the result from vapor-pressure data.  The relationship between the two compressibilities is not entirely clear, but if we 
multiply our result by $k_B$ and divide by the film thickness then we get a value within the same order of magnitude as those earlier results.  

The data points in Fig.\,4(b) show the linear slope with temperature of the third sound velocity in the low-temperature regime well below $T_{KT}$.  The slope oscillates both positive and negative, in phase with the oscillation in the compressibility.  The solid line shows a calculation that assumes that the amplitude of the compressibility of the film decreases linearly with temperature,
\begin{equation}
{K_s}(d,T) = {K_s}(d,0)(1 - \beta T)\quad .
\end{equation}
Such a decrease lowers the maximum third sound speed at 3.0 layers, but then raises it from the minimum value at 3.5 layers, giving rise to the observed positive slope.  To get the observed approximate match it was necessary to take $\beta = 0.1$ K$^{-1}$ and to adjust the third sound index of refraction to 2.0 (the value used for the compressibility of Fig.\,4(a)).  It was also necessary to assume that the temperature dependence of the superfluid density at low temperatures was negligible, which has not been measured for the nanotube geometry.  The linear slope with temperature observed with the the third sound rules out a measurable amplitude for the $T^2$ temperature dependence of the normal fluid component that might be expected for a superfluid with 1D excitations \cite{padmore}.  There could be a linear in T variation that could arise from 0D excitations (i.e. purely azimuthal waves around the nanotube), but this would have to be a very small amplitude, since otherwise the positive low-temperature third sound slope would not be observed.  

\section{Discussion}
  
There have been theoretical studies of third sound  in layered helium films on carbon surfaces \cite{campbell},  which find oscillations in the third sound velocity with layer periodicity.  However, in a major discrepancy with experiments, the velocity is predicted to be zero at layer completion (and for a finite region about the layer completion),  and then a maximum at half-layer thicknesses.  This is just the converse of the behavior seen in our experiments and those of Zimmerli and Chan \cite{zimmerli}, which are a maximum at layer completion and minimum at the half-layer points.  The zero velocity in the theories arises from discontinuities in the chemical potential around layer completions, coming from a first-order ``layering transition" where atoms added to the film are energetically favored to fill in the next higher layer instead of completing the nearly-filled layer, causing the compressibility to become infinite.  It is quite unclear why the experiments do not show any of this predicted behavior.

The linear temperature dependence of the compressibility which we observed even at temperatures near 100 mK was unexpected.  Our original expectation was that at temperatures well below the energies of excitations such as rotons and vortices there would be no temperature dependence at all.  However, one possible source of temperature dependence could come from a thermodynamic relation between the isentropic and isothermal compressibilities:
\begin{equation}
{K_s} = {K_T} - \frac{{{\alpha ^2}}}{{\rho \,c_p}}T
\end{equation}
where $\alpha$ is the expansion coefficient and $c_p$ the specific heat.  For the linear in $T$ behavior to hold for very low temperatures both the isothermal compressibility and the specific heat would need  to be independent of temperature.  Experiments measuring the film heat capacity on a flat Grafoil substrate \cite{greywall} found the expected 2D variation of $T^2$ from surface excitations \cite{padmore}, so for the nanotubes a 0D behavior would be needed.  It is also not clear what the expansion coefficient of the film would be, whether related to the bulk liquid helium value, to the thermal expansion of the free surface interface \cite{campbell,penanen}, or to thickness changes related to changes in the very high pressure gradients in the film.  Measurements of the heat capacity on the nanotubes would be a necessary first step before Eq.\,(6) could be compared with the experimental value of $\beta$ found from the third sound compressibility.

The lack of agreement with the universal KT slope in Fig.\,3 is very puzzling.  There is no question that the theory is correct for 2D superfluids, with validation from extensive theoretical and experimental work \cite{kosterlitzrev}.  We do not believe the nanotube geometry of our measurement makes any difference in this respect, since as seen in Fig.\,3 our results show the same anomalous slope seen on the flat Grafoil substrate \cite{crowell}.

One possible explanation for the disagreement with the 2D KT theory is if the film might in fact have 3D modifications to its structure: ripple-like corrugations on the scale of the substrate carbon-ring structure \cite{mounds}.  Theory has shown that the binding potential of a helium atom is about 20\% stronger at the middle of the carbon rings than at positions right over the carbon atoms \cite{cole}.  This anisotropic potential has been utilized in calculations \cite{manousakis,manousakis1999,troyer,Gordillo2013}, and is seen to result in corrugations of the surface, though the simulations were only carried out in the first and second helium layers. 
It would be interesting if such corrugation effects could be extended to the thicker films we have studied, since the energy to nucleate a vortex is directly proportional to the thickness of the film.  Even a small modulation in thickness could substantially lower the superfluid onset temperature, by allowing an enhanced nucleation of vortices in the thinner regions away from the ring centers.  The observations that did find agreement with the KT slope \cite{rudnick1978,bishop,kotsubo,choprl,cho3} were on highly disordered surfaces such as glass, Mylar, and aluminum oxide, and it may be that the atomic disorder served to average out any corrugation effects, leading to an effectively 2D film.

\section{Conclusions and Outlook}

In summary, we have measured the superfluid onset temperatures as a function of film thickness between 3 and 4.4 atomic layers of $^4$He adsorbed on 10 nm diameter carbon nanotubes.  The onset is linear in the film thickness, but the slope is a factor of three smaller than the universal KT prediction.  This disagreement with theory does not appear to be connected with the curved geometry of the nanotube surface, since analysis of the data in Ref.\,\cite{crowell} on flat carbon (Grafoil) shows the same disagreement.  We speculate that the highly ordered carbon surfaces may give rise to 3D modifications of the film surface, causing enhanced vortex nucleation beyond that predicted by the 2D KT theory.  

We also found evidence of a linear decrease of the isentropic compressibility of the film with temperature, even at very low temperatures where no structural changes in the film were expected.  The reason for this increase in the film stiffness with temperature is not clear.  It is possible it could be related to the thermodynamic difference between the isentropic and isothermal compressibilities, but measurements of the specific heat on the nanotube substrate will be needed to quantify this relation.

Further measurements of helium films on nanotubes will be necessary to understand these results.  Torsion oscillator measurements would allow a direct measurement of the superfluid fraction, something that is difficult with third sound because of the coupling with the compressibility at low temperature and the increasing dissipation at the superfluid onset.  The torsion oscillator would allow a much more complete study of the finite-size aspects of the onset region and comparison with the Machta-Guyer theory \cite{Machta1989}.  It would also allow measurements on thinner films near the re-entrant transition, for comparison with the flat Grafoil measurements \cite{crowell}.  Measurements would also be interesting on nanotubes of smaller diameter; nanotubes can now be fabricated with only 2-3 walls at diameters of order 3 nm.  In this regime the curvature could start to play a major role.

\begin{acknowledgments}
This work was supported in part by a grant from the Julian Schwinger Foundation.  We acknowledge useful discussions with S. Putterman, M. C. Gordillo, and J. Boronat.
\end{acknowledgments}

\bibliography{Nanotube_PRB-3}

\begin{thebibliography}{36}
\bibitem [{mou()}]{mounds}%
  \BibitemOpen
  \href@noop {} {\bibinfo  {journal} {We note that such corrugations could be
  related to the mechanism causing the loss of superfluidity in the second
  layer \cite{nyeki}, which does not show KT behavior. The static surface
  ripples can be thought of as standing waves of third sound with a wavelength
  equal to the spacing between the centers of the carbon rings, 2.46 \AA. These
  could possibly be the soft {``}roton-like" modes with triangular symmetry
  proposed in Ref.[8]}\ }\BibitemShut {NoStop}%
\end{thebibliography}
\end{document}